\documentclass[lettersize,journal]{IEEEtran}
\usepackage{amsmath,amsfonts}
\usepackage{algorithmic}
\usepackage{algorithm}
\usepackage{array}
\usepackage[caption=false,font=normalsize,labelfont=sf,textfont=sf]{subfig}
\usepackage{textcomp}
\usepackage{stfloats}
\usepackage{url}
\usepackage{verbatim}
\usepackage{graphicx}
\usepackage{cite}
\usepackage{booktabs}
\usepackage{multirow}
\usepackage[hidelinks]{hyperref}
\usepackage{cleveref}
\usepackage{nicefrac}
\usepackage{bm}
\usepackage{mathtools}
\usepackage{enumitem}
\usepackage{siunitx}
\allowdisplaybreaks

\input{acro}

\begin{document}

\title{Implementing General-Order Frequency Dynamic Response Model and Frequency Excursion Duration Criterion in Unit Commitment Problem}

\author{Mohammad Rajabdorri, Bo Zhou,~\IEEEmembership{Member,~IEEE,} Lukas Sigrist,~\IEEEmembership{Member,~IEEE,} Enrique Lobato
\thanks{M. Rajabdorri, E. Lobato, and L. Sigrist are with the Instituto de Investigación Tecnológica (IIT), Universidad Pontificia Comillas, Madrid, Spain (e-mail: \{mrajabdorri, enrique, lsigrist\}@comillas.edu).}

\thanks{B. Zhou is Bo Zhou is a research fellow in the Department of Industrial and Operations Engineering at the University of Michigan. (e-mail: bozum@umich.edu).}
}
\maketitle

\begin{abstract}
This paper introduces a novel approach for incorporating frequency dynamics into the unit commitment (UC) problem through a general-order differential equation model, solved using Bernstein polynomial approximation. Traditional frequency-constrained UC (FCUC) models typically rely on simplified first-order assumptions or scalar frequency metrics, such as frequency nadir, to indirectly enforce dynamic behavior. In contrast, our formulation explicitly models time-domain frequency response using second-order dynamics, enabling a more accurate and flexible representation of generator behavior. The resulting differential equations are approximated with high fidelity using Bernstein polynomials, leading to a mixed-integer linear programming (MILP) formulation that remains computationally tractable for small-scale power systems.

Additionally, we introduce a new constraint based on the duration of frequency excursions below a critical threshold, motivated by practical concerns such as relay operation and equipment protection. A data-driven method is employed to relate the area under this threshold—computed as the integral of the Bernstein approximation—to the duration of frequency deviation. The proposed framework is validated using real-world data from an island system in Spain, demonstrating enhanced frequency security with a moderate increase in operational cost. These results suggest the method’s strong potential for application in low-inertia, small-scale power systems.
\end{abstract}



\section{Introduction}

\subsection{Background and Motivation}

Maintaining power system stability is becoming increasingly challenging as traditional synchronous generators are progressively replaced by non-synchronous renewable energy sources. This challenge is especially critical in smaller power systems, where limited inertia significantly increases the risk of instability. For instance, in July 2020, the Spanish island of Tenerife experienced a complete blackout triggered by the shutdown of a power plant. The initial outage led to cascading failures and severe frequency deviations. Although the \gls{UFLS} scheme was activated, demand reduction was not sufficient, ultimately resulting in a total blackout that lasted approximately seven hours \cite{REE2020Tenerife}.

To prevent such incidents in the operational stage, several preventive approaches have been proposed in the literature by integrating frequency-related constraints into scheduling problems, such as unit commitment. These methods aim to capture the frequency dynamics resulting from specific outage scenarios in order to avoid unacceptable frequency deviations \cite{badesa2019simultaneous,trovato2018unit,jiang2022coordinative,han2024robust,zhou2023frequency,lagos2021data,zhang2020approximating,sang2023conservative,rajabdorri2022robust,rajabdorri2023inclusion,rajabdorri2024data}.
Unlike traditional power system scheduling problems, typically solved with an hourly resolution, frequency response unfolds on a much faster timescale, often reaching critical thresholds within seconds. Moreover, these two domains differ significantly in mathematical formulation: scheduling problems are usually NP-hard and are often modeled as \gls{MILP}, whereas frequency dynamics are described using continuous-time \glspl{ODE}. These differences present significant challenges when attempting to incorporate frequency dynamics into conventional scheduling frameworks.

The most common approach to incorporating frequency dynamics into power system scheduling is to extract key frequency metrics from simplified dynamic models and embed them into the optimization formulation \cite{badesa2019simultaneous,trovato2018unit,jiang2022coordinative,han2024robust}. Typical metrics include \gls{RoCoF}, frequency nadir, and quasi-steady-state frequency.
However, the effectiveness of these metrics is subject to debate. In practice, most equipment is sensitive not only to the magnitude of frequency deviations but also to the duration for which the frequency remains below a critical threshold \cite{tierney2014performance}. Protective relays, for instance, are typically configured to respond based on the duration of such deviations according to their timeout settings \cite{urban2021estimating}.
Both analytical and data-driven techniques have been widely explored in the literature to estimate these frequency metrics.

Analytical methods often rely on simplifying the well-known swing equation so that the aforementioned frequency metrics become mathematically tractable and can be directly incorporated into \gls{MILP}-based scheduling problems. Among these metrics, the frequency nadir is particularly challenging to compute accurately. A common simplification assumes that generators respond linearly, delivering their reserves within a fixed time frame—typically set to 10 seconds. However, this assumption is questionable, as generator dynamics are more accurately described by second- or even third-order differential equations, depending on the technology and associated time constants \cite{huang2020generic}.

To address this complexity and more accurately capture the reserve response of generators, data-driven methods have been proposed \cite{lagos2021data,zhang2020approximating,sang2023conservative,rajabdorri2022robust,rajabdorri2023inclusion,rajabdorri2024data}. These approaches utilize detailed frequency response simulations to generate labeled datasets, from which machine learning models are trained to estimate frequency metrics. A key advantage of data-driven techniques lies in the accuracy of the labels, as the frequency metrics are directly derived from high-fidelity simulation results. However, a potential limitation is the inevitable loss of precision introduced by the learning algorithm and the quality of the training dataset.

Bernstein polynomials form a basis for the space of continuous functions over a finite interval \cite{lorentz2012bernstein}. They allow for the accurate approximation of solutions to \gls{ODE} problems while maintaining relatively low computational complexity \cite{bhatti2007solutions}. In the context of power system operation and scheduling, Bernstein polynomials are particularly attractive due to their compatibility with \gls{MILP} formulations \cite{parvania2015unit,rajabdorri2024data}. For example, \cite{zhou2023frequency} applies Bernstein polynomials to model frequency response dynamics within the \gls{UC} framework. These and related studies are briefly reviewed in the following section.

\subsection{Literature Review}

\Cref{tab:nadir_modeling} summarizes recent contributions in the context of frequency nadir modeling within unit commitment. Of particular relevance to our work are the techniques used to approximate or predict frequency nadir and the assumptions to do so.

\begin{table*}[!t]
\caption{Summary of Frequency Nadir Modeling in the Literature}
\label{tab:nadir_modeling}
\centering
\renewcommand{\arraystretch}{1.2}
\setlength{\tabcolsep}{3pt}
\small
\begin{tabular}{>{\centering\arraybackslash}m{0.7cm}|
                >{\centering\arraybackslash}m{7.2cm}|
                >{\centering\arraybackslash}m{9.5cm}}
\hline
\textbf{Ref.} & \textbf{Frequency Nadir Modeling} & \textbf{Underlying Assumptions} \\
\hline
\cite{badesa2019simultaneous} & Inner approximation combined with binary expansion. & Derived from the swing equation assuming linear primary response over 10 seconds, including load damping effects. \\
\cite{trovato2018unit} & Piecewise linear upper and lower bounds. & Based on the swing equation with a 10-second linear primary response, capturing fast dynamics of frequency-supporting resources. \\
\cite{jiang2022coordinative} & Second-order conic reformulation of the nadir term. & Derived from the swing equation, neglecting load damping effects. \\
\cite{han2024robust} & Piecewise linear representation. & Incorporates linear primary frequency responses from batteries, thermal units, and wind turbines using the swing equation. \\
\cite{zhou2023frequency} & Inner approximation of differential equations via Bernstein polynomials. & Based on the swing equation with first-order differential-algebraic modeling of generator responses. \\
\cite{lagos2021data} & Constraint generation using optimal classification trees. & Labeled using simulated trajectories from the swing equation with linear primary response. \\
\cite{zhang2020approximating} & Deep neural network prediction embedded in MILP. & Data generated via positive-sequence power balance, labeled using a first-order frequency model. \\
\cite{sang2023conservative} & Conservative sparse neural network predictor. & Dataset generated using boundary and Latin hypercube sampling, labeled via the swing equation with first-order dynamics. \\
\cite{rajabdorri2022robust} & Logistic regression classifier. & Labels derived from simulated frequency response using a second-order model. \\
\cite{rajabdorri2023inclusion} & Logistic regression and support vector machine ensemble. & Frequency nadir labeled from a synthetically generated dataset using a second-order dynamic model. \\
\cite{rajabdorri2024data} & Logistic regression on Bernstein polynomial coefficients. & Labels based on a second-order frequency response model approximated via Bernstein polynomials. \\
\hline
\end{tabular}
\end{table*}

\subsection{Gaps, Contributions, and Paper Outline}

We argue that the common assumption of linear reserve delivery within a fixed time frame—widely adopted in analytical \gls{FCUC} models, does not accurately capture generator dynamics. While this simplification may be acceptable in large-scale systems where frequency deviations are typically modest, it becomes problematic in small or islanded power systems, where frequency can drop sharply following an outage. In such systems, the assumption must be revisited. Although more accurate models have been proposed, such as the first-order dynamic formulation in \cite{zhou2023frequency}, these approaches still simplify generator behavior and leave room for further refinement.

Another important limitation in existing literature is the frequent omission of generator capacity constraints in frequency response modeling. In practice, the maximum deliverable power of each generator is bounded by its nominal capacity, and neglecting this constraint can result in physically infeasible solutions. While incorporating such limits increases the complexity of the \gls{MILP} formulation, it is essential for realism and reliability.

Furthermore, most prior \gls{FCUC} models rely exclusively on frequency nadir as the primary security metric, neglecting the duration for which the frequency remains below a critical threshold. This omission is significant, as prolonged frequency deviations pose a greater risk to equipment and system integrity than short, deep dips. In practice, protection relays and equipment response are often governed by the duration of frequency excursions rather than their absolute nadir. To address this, we propose a novel method that explicitly models and constrains the duration of frequency deviation below threshold values.
While our proposed method introduces greater computational complexity, it remains tractable for small-scale systems, such as islanded grids, where high-fidelity frequency modeling is most needed. Larger systems, which benefit from greater inertia and stability, typically do not require this level of frequency-detail modeling.

This paper introduces a modeling framework that incorporates general-order generator dynamics into the \gls{UC} problem via Bernstein polynomial approximation. It further proposes a data-driven method to constrain the duration frequency under a predefined threshold. To the best of our knowledge, this work introduces several novel contributions:
\begin{itemize}
\item A \gls{MILP}-compatible solution to the swing equation using a general-order dynamic model based on Bernstein polynomials, extending the first-order formulation of \cite{zhou2023frequency}.
\item Explicit inclusion and linearization of generator capacity constraints within the dynamic frequency response model.
\item A new constraint that limits the \textit{duration} for which system frequency remains below a critical threshold—rather than constraining only the nadir.
\item A data-driven approach for estimating the duration of frequency excursions using the integral of the Bernstein polynomial approximation.
\item Validation on a real-world islanded power system in Spain, demonstrating the model’s effectiveness in enhancing frequency security.
\end{itemize}

The remainder of this paper is organized as follows. \Cref{sec:generalOrder} introduces Bernstein polynomial theory and its application to differential equations. \Cref{sec:dataDrivenApproach} presents the data-driven method for estimating frequency excursion duration. In \cref{sec:implementation}, we integrate the proposed constraints into the \gls{UC} problem. Numerical results are presented and discussed in \cref{sec:result}, followed by conclusions in \cref{sec:conc}.

\section{Solving General-Order Frequency Response with Bernstein Polynomials}\label{sec:generalOrder}

\subsection{Bernstein Polynomials}

\subsubsection{Polynomial Basis} The Bernstein polynomial of a function \(g(t)\), defined over the closed interval \(t \in [0, h]\), is computed as a weighted sum of Bernstein basis functions. Specifically, \({B_n}^g(t)\) denotes the Bernstein polynomial of order \(n\) corresponding to \(g(t)\), and is given by (see \cite{dierckx1995curve}),
\begin{equation}\label{eq:ber_poly}
    {B_n}^g(t)=\sum_{j=0}^{n}{C_j}^g\times \beta_{j,n}(t),
\end{equation}
Where ${C_j}^g$ is the Bernstein coefficient and for $g(t)$ is equal to $g\big(\nicefrac{j}{n}\big)$. ${B_n}^g(t)$ is a polynomial in $t$ of order $n$ and $\beta_{j,n}(t)$ are the Bernstein basis functions.
\begin{equation}
    \beta_{j,n}(t)=\binom{n}{j}\frac{t^j(h-t)^{n-j}}{h^n}
\end{equation}

For the purposes of this work, it is more convenient to express Bernstein polynomials in vector form. The vector form will be utilized consistently throughout the paper.
\begin{equation}\label{eq:bernstein}
    g(t)\approx{B_n}^g(t)=
        {\begin{bmatrix}
        \beta_{0,n}(t)\\ \beta_{1,n}(t) \\ \vdots \\ \beta_{n,n}(t)
    \end{bmatrix}}^\top
    \begin{bmatrix}
        {C_{0}}^g \\ {C_{1}}^g \\ \vdots \\ {C_{n}}^g
    \end{bmatrix}
    = \bm{\beta}_n\bm{G}
\end{equation}
\Cref{eq:bernstein} represents the Bernstein polynomials of order $n$ approximating function $g(t)$ defined on the closed interval $[0,h]$, with Bernstein coefficients, ${C_{0}}^g$ to ${C_{n}}^g$, and Bernstein basis functions, $\beta_{0,n}(t)$ to $\beta_{n,n}(t)$. $\bm{\beta}_n$ is the vector of Bernstein basis functions, and $\bm{G}$ is the vector containing the Bernstein coefficients of the function $g(t)$.

\subsubsection{Matrix of Integration}
Bernstein polynomials of the same order can be easily summed by adding their corresponding coefficients. To preserve the polynomial order during integration or differentiation, operational matrices are introduced in \cite{yousefi2010operational}. Given the definition of the Bernstein approximation in \cref{eq:bernstein}, the integral of ${B_n}^g(t)$ is calculated as follows,
\begin{equation}\label{eq:integral}
    \int_0^t{B_n}^g(t)dt \approx \bm{\beta}_nX_n^\top\bm{G}
\end{equation}
Where $X_n$ is the $(n+1)\times(n+1)$ operational matrix of integration for a Bernstein polynomial of order $n$.
\subsubsection{Vector of variable}
A monomial term $t^i$ can be represented in the Bernstein basis of order \(n\) as,
\begin{equation}
    t^i=\beta_n
    \begin{bmatrix}
        0 \\ (\frac{1}{n})^i \\ \vdots \\ (\frac{n-1}{n})^i \\ 1
    \end{bmatrix}
    =\beta_n\bm{d}_{i,n}
\end{equation}
Here, \(\bm{d}_{i,n}\) is a \((n+1) \times 1\) vector containing the evaluations of \(t^i\) at uniformly spaced Bernstein nodes. This vector will be used later to express derivatives within the Bernstein polynomial framework.

\subsubsection{Solving Differential Equations with Bernstein Polynomials}\label{sec:solving_ODE}

A general form of an ordinary differential equation of order $s$ can be defined as \cite{ordokhani2011approximate},
\begin{subequations}\label{eq:ODE}
    \begin{align}
    &\sum_{j=0}^{s} \alpha_j(t)y^{(j)}(t)=f(t), && 0<t<h
    \end{align}
\end{subequations}
Where $\alpha_j(t)$ and $f(t)$ are either known functions or constants. It's assumed that the initial condition of the derivatives, $y^{(j)}(0)$, are known. To solve the \cref{eq:ODE} the functions are substituted with their Bernstein approximation, as previously shown in \cref{eq:ber_poly}. The idea is to express all derivatives in terms of the highest-order derivative and the known initial conditions. We approximate $y^{(s)}(t)$ to a Bernstein polynomial of order $n$ in the interval of $[0,h]$,
\begin{equation}
    y^{(s)}(t) \approx \beta_n \bm{Y}^{(s)}
\end{equation}
Then the other derivatives are defined as\cref{eq:derivatives}.
\begin{figure*}[!t]
\begin{equation}\label{eq:derivatives}
    \begin{gathered}
        \bm{Y}^{(s-1)}=y^{(s-1)}(0)+hX^\top\bm{Y}^{(s)}\\
        \bm{Y}^{(s-2)}=y^{(s-2)}(0)+hy^{(s-1)}(0)\bm{d}_1+h^2(X^\top)^2\bm{Y}^{(s)}\\
        \vdots\\
        \bm{Y}=y^{(0)}(0)+hy^{(1)}(0)\bm{d}_1+h^2\frac{y^{(2)}(0)}{2!}\bm{d}_2+\cdots+h^{s-1}\frac{y^{(s-1)}(0)}{(s-1)!}\bm{d}_{s-1} + h^s(X^\top)^s\bm{Y}^{(s)}
    \end{gathered}
\end{equation}
\end{figure*}
By using \cref{eq:derivatives}, all the derivatives in \cref{eq:ODE} will be written as a function of initial conditions and the Bernstein coefficients of highest order derivative, $\bm{Y}^{(s)}$, and \cref{eq:ODE} can be solved as a system of equations.j

\subsection{First Order Model}\label{sec:FOM}

The first-order differential equations governing frequency response can be written as \cite{sauer2017power},
\begin{subequations}\label{eq:first_order_SFR}
\begin{align}
    2H^{\text{\tiny sys}}\frac{d\Delta f(t)}{dt}+DP_d\Delta f(t) &= p_\ell -\sum_{i}\min\big( r_i(t),\rho_i\big) 
    \label{eq:swing}\\
    T_i\frac{dr_i(t)}{dt}+r_i(t)=K_ix_i&(\Delta f(t)+b_i\frac{d\Delta f(t)}{dt})\label{eq:first_order_r}
\end{align}
\end{subequations}
Where $H^{\text{sys}}$ is the equivalent system inertia in seconds, $D$ is the load damping factor (in \%), $P_d$ is the demand (in per unit), $\Delta f(t)$ is the frequency changes (in per unit), $p_\ell$ is the power mismatch (in per unit), $r_i(t)$ is the reserve provided by unit $i$ (in per unit), $\rho_i$ is the available reserve of unit $i$ and hence the maximum value that $r_i(t)$ can reach,
$T_i$ is the response constant of governors, $K$ is the turbine-governor gain, and $b$ is the multiplier of frequency derivative. $x$ is the commitment variable of the unit. The \texttt{min(.)} function that is used in the swing equation is to limit the responded reserve of units ($r_i(t)$) to their available reserve ($\rho_i$). A discussion on the relevance of this \texttt{min(.)} function is presented in appendix I.

\begin{figure}[!htbp]
    \centering
    \includegraphics[width=0.65\linewidth]{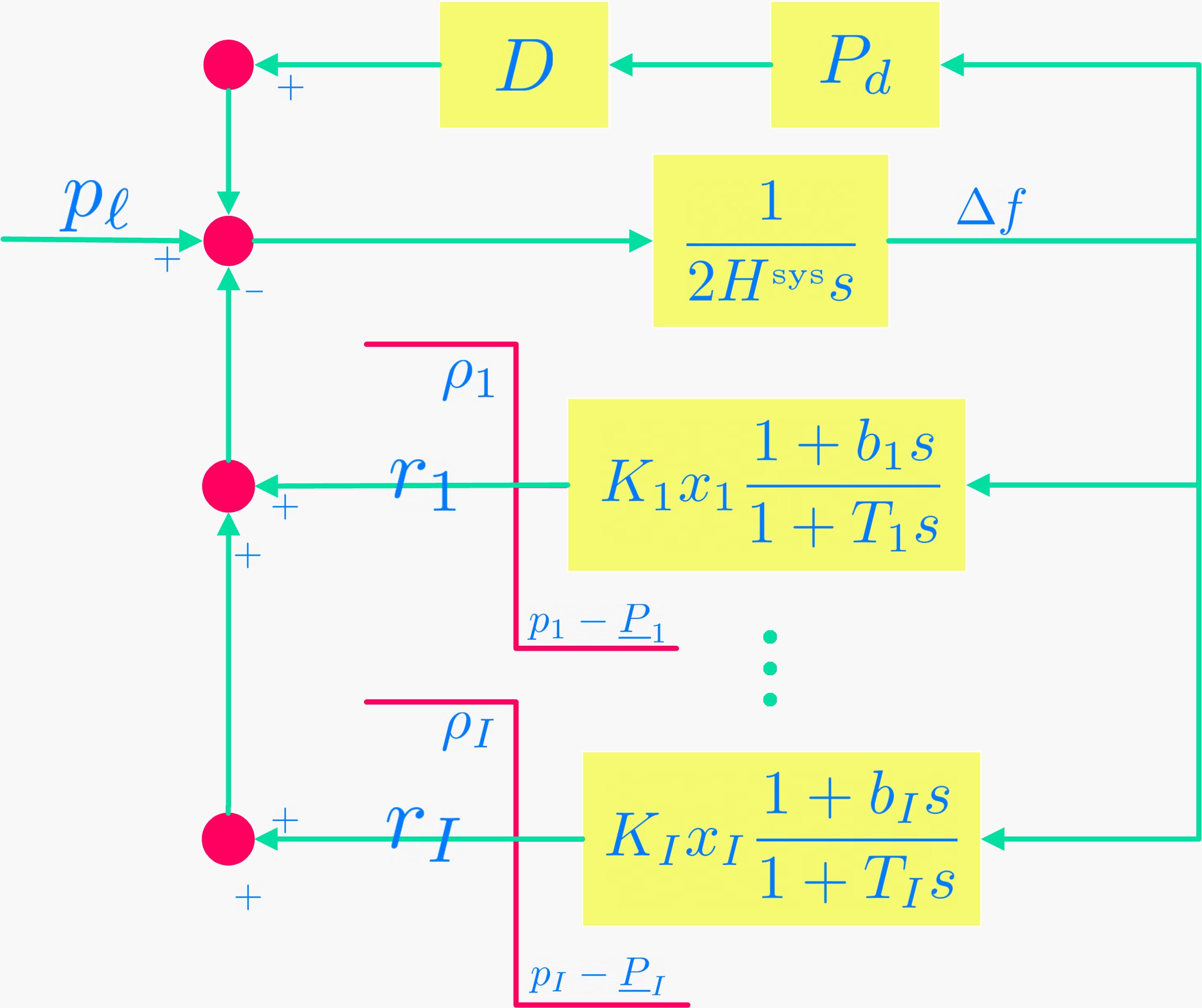}
    \caption{First order model schematic}
    \label{fig:firstOrderModelSchematic}
\end{figure}

The formulation in \cref{eq:first_order_SFR} comprises the swing equation in \cref{eq:swing}, the first-order differential equation of power reserve for each unit in \cref{eq:first_order_r}. 
Additional resources such as \glspl{BESS}, \glspl{RES}, flywheels with appropriate controls, and any other elements offering primary frequency response can be modeled and seamlessly integrated into the formulation as needed.
The model's schematic representation in Laplace form is depicted in \cref{fig:firstOrderModelSchematic}.
\cite{zhou2023frequency} utilizes Bernstein polynomials to approximate the solution of a comparable model. The key distinction lies in the Laplace transfer function, which in \cite{zhou2023frequency}, has no zeros. As outlined in \cite{zhou2023frequency}, the differential equation in \cref{eq:first_order_SFR} can be approximated effectively using Bernstein polynomials.

Both sides of the differential equations in \cref{eq:first_order_SFR} are integrated. Leveraging the operational matrices derived in \cite{yousefi2010operational}, the differential equations are transformed into algebraic equations. We define vectors of Bernstein coefficients for frequency deviation ($\bm{F}$) and power reserve ($\bm{R}$) as follows,
\begin{equation}
    \bm{F}=
    \begin{bmatrix}
        {C_{0}}^{\Delta f} \\ {C_{1}}^{\Delta f} \\ \vdots \\ {C_{n}}^{\Delta f}
    \end{bmatrix},\;\;\;\;
    \bm{R}=
    \begin{bmatrix}
        {C_{0}}^{r} \\ {C_{1}}^{r} \\ \vdots \\ {C_{n}}^{r}
    \end{bmatrix}    
\end{equation}

Here is a summary of the Bernstein approximation of \cref{eq:first_order_SFR}.
\begin{subequations}\label{eq:first_order_BP}
    \begin{align}
    &\begin{aligned}
        \frac{2H^{\text{\tiny sys}}}{h_\tau}\big(\bm{F}_\tau-\Delta f_{\tau}^0\big)+DP_dX^\top\bm{F}_\tau & = \\  X^\top\big(p_\ell -\sum_{i}&\min(\bm{R}_{\tau,i},\rho_i)
        \big) \label{eq:swing_br}
    \end{aligned}\\
    &\begin{aligned}
        \frac{T_i}{h_\tau}\big(\bm{R}_{\tau,i}-r_{\tau,i}^0\big)+X^\top\bm{R}_{\tau,i} & = \\ K_ix_i\big(X^\top\bm{F}_\tau&+\frac{b_i}{h_\tau}(\bm{F}_\tau-\Delta f_{\tau}^0)\big) 
    \end{aligned}\label{eq:first_order_r_br}
    \end{align}
\end{subequations}
Where $h_\tau$ is the length of the time segment. A discussion on choosing $h_\tau$ for the purpose of approximating frequency response is presented in appendix II. $\bm{F}$ is the vector of Bernstein coefficients of frequency. For Bernstein polynomials of order $n$, $\bm{F}$ will be a $(n+1)\times 1$ vector. A discussion on choosing $n$ is presented in appendix III. $\tau$ is the index of the time segment. $X^\top$ is the transpose of the operational matrix of integration in \cref{eq:integral}. $p_\ell$ is the lost power. $\bm{R}$ 
is the vectors of Bernstein coefficients of the power reserve of the unit.
The initial vectors of frequency and power reserve ($\Delta f_{\tau}^0$, $r_{\tau}^0$) are obtained from the solutions of the previous segment. For example, $\Delta f_{\tau}^0$ is the last element of $\bm{F}_{\tau-1}$.
Solving \cref{eq:first_order_BP} for consecutive time segments approximates the first-order differential equations of the dynamic response. The algebraic equations in \cref{eq:first_order_BP} are bilinear and can be linearized for implementation in the \gls{UC} problem.

While the first-order model captures essential system dynamics, it may not adequately represent faster movements. Research by \cite{huang2020generic} indicates that a first-order model for mover-governor systems lacks accuracy. Instead, a second-order transfer function for the equivalent prime mover-governor is recommended. In the following sections, we introduce a second-order model as a representative case, although the same Bernstein approximation framework is directly applicable to models of any order.

\subsection{Second Order Model}\label{sec:SOM}

\begin{figure}[!htbp]
    \centering
    \includegraphics[width=0.65\linewidth]{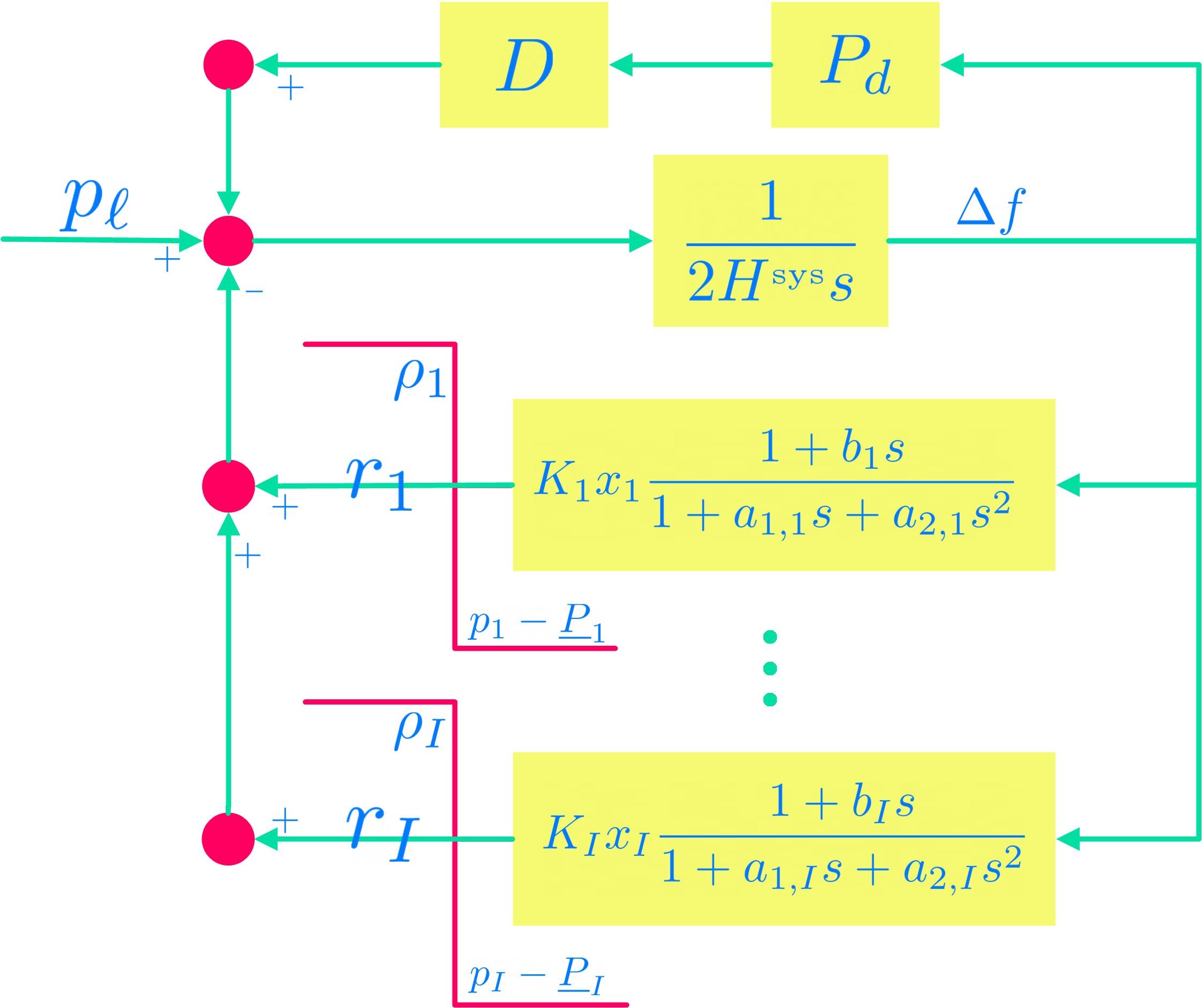}
    \caption{Second order model schematic}
    \label{fig:secondOrderModelSchematic}
\end{figure}

A second-order model representing the schematic of \cref{fig:secondOrderModelSchematic} is presented here. How a second-order model response can differ from the first order is discussed in appendix IV.

\begin{subequations}\label{eq:second_order}
    \begin{align}
            2H^{\text{\tiny sys}}\frac{d\Delta f(t)}{dt}+DP_d\Delta f(t) &= p_\ell -\sum_{i}\min \big(r_i(t),\rho_i\big) 
            \label{eq:swing2}\\
            a_{2,i}\frac{d^2r_i}{dt^2}+a_{1,i}\frac{dr_i}{dt}+r_i(t) &= K_ix_i(\Delta f(t)+b_i\frac{d\Delta f(t)}{dt})\label{eq:second_order_r}
    \end{align}
\end{subequations}

Building upon the groundwork established in \cite{zhou2023frequency}, we derive an approximate solution to \cref{eq:second_order} using the general method that is presented in \cref{sec:solving_ODE}. Here, we outline the general form of the formulation, with $h_\tau$ denoting the length of time segments and $n_\tau$ representing the order of the Bernstein polynomial for segment $\tau$.
Initially, we approximate $\frac{d^2r_i}{dt^2}$ and $\frac{d\Delta f(t)}{dt}$, which are the highest order derivatives of each variable, with Bernstein polynomials of order $n_\tau$ in each segment $\tau$,
\begin{equation}\label{eq:r2ber}
    \frac{d^2r_i}{dt^2} \approx \beta_{n_\tau} \bm{R}^{(2)}_{\tau,i}
\end{equation}
\begin{equation}\label{eq:f1ber}
    \frac{d\Delta f(t)}{dt} \approx \beta_{n_\tau} \bm{F}^{(1)}_{\tau}
\end{equation}
Using the approximations in \cref{eq:r2ber} and \cref{eq:f1ber} we define the vector of Bernstein coefficients for the rest,
\begin{equation}\label{eq:r1ber}
    \bm{R}^{(1)}_{\tau,i}=r^{(1)}_{\tau,i}(0)+h_\tau X_{n_\tau}^\top\bm{R}^{(2)}_{\tau,i}
\end{equation}
\begin{equation}\label{eq:r0ber}
    \bm{R}_{\tau,i}=r^{(0)}_{\tau,i}(0) + h_\tau r^{(1)}_{\tau,i}(0)\bm{d}_{1,n_\tau}
    + h_\tau^2 (X_{n_\tau}^\top)^2\bm{R}^{(2)}_{\tau,i}
\end{equation}
\begin{equation}\label{eq:f0ber}
    \bm{F}_{\tau}=\Delta f^{(0)}_{\tau}(0)+h_\tau X_{n_\tau}^\top\bm{F}^{(1)}_{\tau}
\end{equation}
The initial conditions of the derivatives is zero for the first segment and for the other segments is obtained from the solution of the previous segment. For example, $r(0)^{(1)}_{\tau,i}$ is the last element of $\bm{R}^{(1)}_{i,\tau-1}$.

Now we rewrite \cref{eq:swing2,eq:second_order_r} with the terms that are defined in \cref{eq:r2ber,eq:f1ber,eq:r1ber,eq:r0ber,eq:f0ber},
\begin{subequations}\label{eq:second_order_ber}
    \begin{align}
    2H^{\text{\tiny sys}}\bm{F}^{(1)}_{\tau}+DP_d\bm{F}_{\tau}  = p_\ell -&\sum_{i}\min \big(\bm{R}_{\tau,i},\rho_i\big) 
    \label{eq:swing2_ber}\\
    a_{2,i}\bm{R}^{(2)}_{\tau,i} +a_{1,i}\bm{R}^{(1)}_{\tau,i}+\bm{R}_{\tau,i} &= K_ix_i(\bm{F}_{\tau}+b_i \bm{F}^{(1)}_{\tau})\label{eq:second_order_r_ber}
    \end{align}
\end{subequations}
\Cref{eq:swing2_ber,eq:second_order_r_ber} are algebraic systems of equations and straight forward to solve. There are some nonlinearities, namely the binary into variables in $H^{\text{\tiny sys}}\bm{F}^{(1)}_{\tau}$, $x_i\bm{F}$, and $x_i\bm{F}^{(1)}_{\tau}$, and the \texttt{min(.)} function. The linearization of binary into variable and \texttt{min(.)} are respectively explained in appendices V and VI.  

\section{Frequency Excursion  Duration Constraint}\label{sec:dataDrivenApproach}

The frequency nadir is a key metric for evaluating system frequency response. Based on the derivations presented in \cref{sec:generalOrder}, a frequency nadir constraint can be readily formulated. Specifically, this is achieved by ensuring that all Bernstein polynomial coefficients representing the frequency trajectory remain above the specified nadir threshold.
\begin{equation}\label{eq:nadir}
    \bm{F}\geq \Delta f_{\text{th}}
\end{equation}
While numerous prior studies impose frequency nadir constraints \cite{badesa2019simultaneous,trovato2018unit,jiang2022coordinative,han2024robust
,zhou2023frequency,lagos2021data,zhang2020approximating,sang2023conservative,rajabdorri2022robust,rajabdorri2023inclusion,rajabdorri2024data}
, it's noteworthy that \gls{UFLS} schemes and relays are often triggered based on the duration that frequency remains below a threshold, rather than the minimum value that frequency reaches.

\begin{figure}[!htbp]
    \centering
    \includegraphics[width=0.9\linewidth]{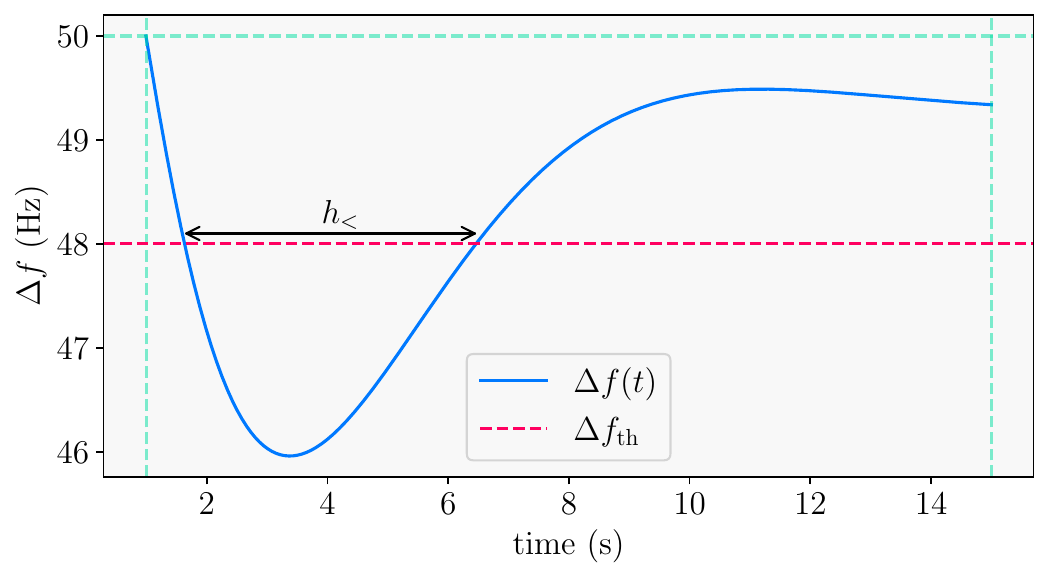}
    \caption{Frequency response after outage}
    \label{fig:freq_response}
\end{figure}

\begin{figure}[!htbp]
    \centering
    \includegraphics[width=0.9\linewidth]{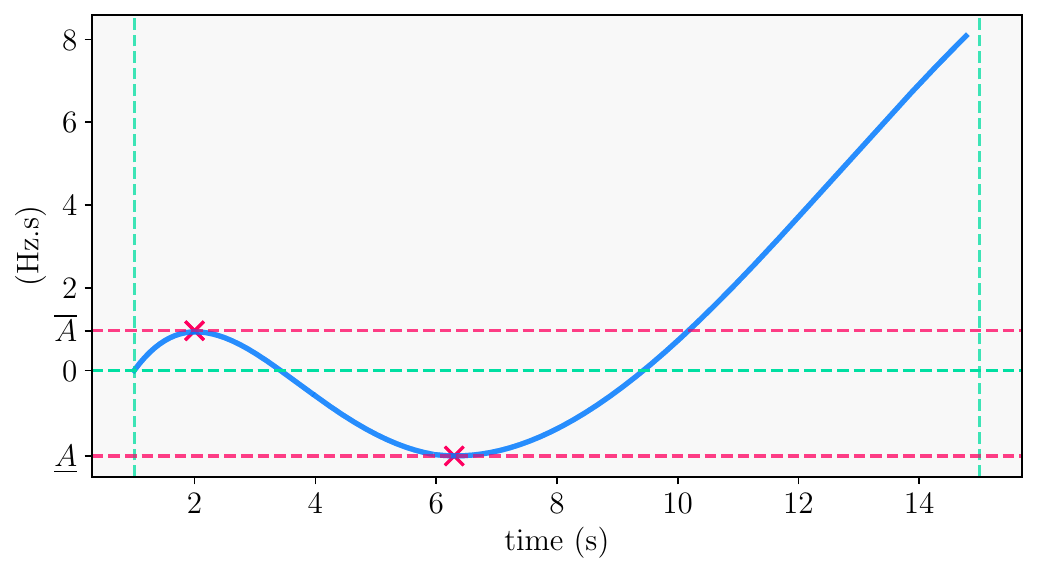}
    \caption{Integral of $\Delta f(t)-\Delta f_{\text{th}}$}
    \label{fig:integ_freq_response}
\end{figure}

While by assigning binaries to each Bernstein coefficient, we can approximate the duration of frequency decline, introducing so many binaries is computationally heavy. On the other hand, as it's stated before in \cref{eq:integral}, the integral of a Bernstein polynomial can be accurately estimated just by multiplying it in a matrix with known values. The integral of $\Delta f(t)-\Delta f_{\text{th}}$ contains some valuable information about the duration of frequency decline and the severity of frequency drop. Frequency response after an outage looks something like \cref{fig:freq_response}.
Frequency might pass the threshold ($\Delta f_{\text{th}}$) and then recover. We're interested in limiting the duration frequency under $\Delta f_{\text{th}}$. By doing so, we can allow incidents after which frequency goes below the threshold, but returns above the threshold shortly after. It's intuitive to assume the area below the threshold is bigger, probably the duration of the frequency under the threshold is longer. Here the integral of $\Delta f(t)-\Delta f_{\text{th}}$ might have some useful information. Integral of $\Delta f(t)-\Delta f_{\text{th}}$ for a frequency response similar to \cref{fig:freq_response} is something like \cref{fig:integ_freq_response}.
The value of the integral will increase until the frequency reaches to $\Delta f_{\text{th}}$ (a local maximum value), then it will start going down until it reaches $\Delta f_{\text{th}}$ again (a minimum value), and then it starts increasing again. This is for the outages that are big enough that the frequency goes below the threshold, and outages small enough that the frequency will recover at some point (reaches the threshold again). If the outage is small and frequency never reaches the threshold, the integral always increases. On the other hand, if the outage is very big and the frequency keeps going down, the integral will always go down after reaching a maximum. We define,
\begin{equation}\label{eq:a_integral}
    A=\int_0^t [\Delta f(t)-\Delta f_{\text{th}}]dt
\end{equation}
Note that if the frequency is defined as a Bernstein polynomial, \cref{eq:a_integral} is calculated as,
\begin{equation}\label{eq:area_ber}
    \bm{A}\approx \bm{\beta}_nX_n^\top(\bm{F}-\Delta f_{\text{th}})
\end{equation}
To calculate the area under threshold (in \cref{fig:freq_response}), we need to calculate the first local maximum value ($\overline{A}$) and the minimum value ($\underline{A}$) of the integral, and the aforementioned area would be $\overline{A}-\underline{A}$ (look at \cref{fig:integ_freq_response}). There are two obstacles here; First, finding $\overline{A}$ and $\underline{A}$, although possible, is not straightforward. Second, the calculated area wouldn't directly give us the duration of the frequency being under the threshold. We are interested in preventing incidents, in which the frequency goes below the threshold for longer than $h_{\text{th}}$. $\underline{A}$ is a good indicator for that. Instinctively, it seems that for lower $\underline{A}$, the duration of frequency under threshold is probably higher. This claim should be further studied.

\begin{figure}[!htbp]
    \centering
    \includegraphics[width=1\linewidth]{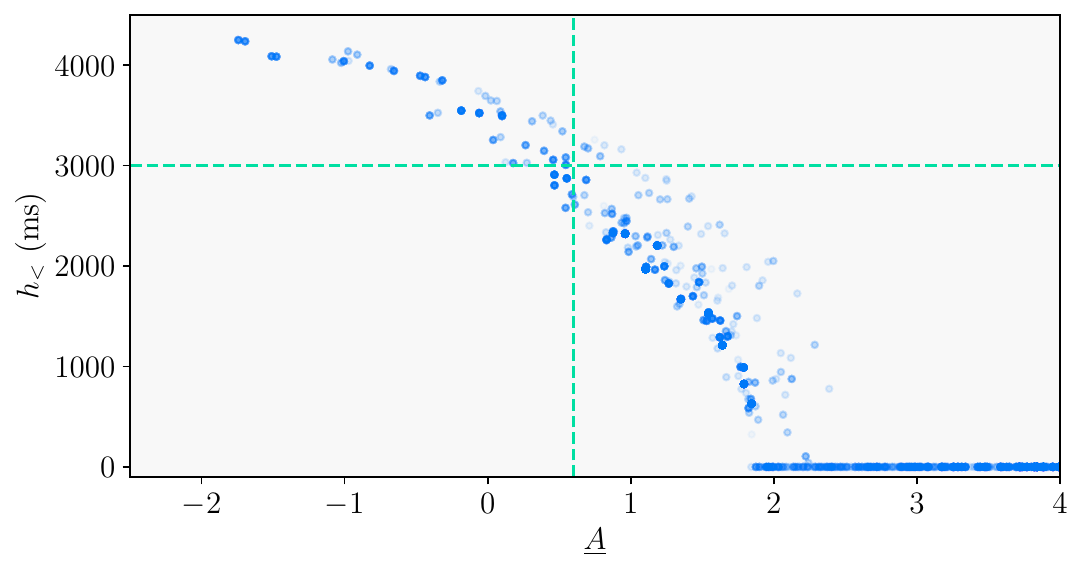}
    \caption{The duration of frequency under threshold ($h_<$) as a function of $\underline{A}$.}
    \label{fig:durationVsArea}
\end{figure}

We conducted a data-driven analysis to see if $\underline{A}$ is a good indicator of the duration of frequency under threshold or not. The dataset obtained from the synthetic data generation method, introduced by \cite{rajabdorri2023inclusion}, is utilized. Frequency response after more than 100,000 outages are is calculated using \cref{eq:second_order}. \Cref{fig:durationVsArea} shows the obtained result.
As shown in \cref{fig:durationVsArea}, a larger value of $\underline{A}$ results in a shorter $h_<$. In cases where the frequency response does not reach the threshold, $A$ remains non-negative (see \cref{eq:integral}), hence $h_<=0$. When the frequency response exceeds the threshold, $A$ reaches a local minimum, and $\underline{A}$ depends on the area beneath the threshold and the frequency curve (see \cref{fig:freq_response}). We propose that by imposing a constraint on $\underline{A}$, we can control $h_<$, which is the primary goal of this paper. This approach offers a more direct method from a modeling perspective. $A$ can be computed using \cref{eq:area_ber}, and by constraining the control points of $A$, we can effectively regulate $\underline{A}$.
\begin{equation}\label{eq:a_constraint}
    \bm{A}\geq A_{\text{th}}
\end{equation}

\section{Implementation in UC}\label{sec:implementation}


To test the effectiveness of the proposed solution for the second-order differential model of the dynamic response, we implement it in the \gls{UC} problem. A general \gls{UC} formulation is presented here,
\begin{subequations}\label{eqs:uc}
\begin{align}
\min_{x,p}&\ \text{suc}(x)+\text{gc}(p) \label{eq:ucObj}
\\
x_{t,i}-x_{t-1,i}&=y_{t,i}-z_{t,i} &&\text{\scriptsize $t\in \mathcal{T},\;i\in\mathcal{I}$}\label{eq:ucBin1}
\\
y_{t,i}+z_{t,i}&\leq1 &&\text{\scriptsize $t\in \mathcal{T},\;i\in\mathcal{I}$}\label{eq:ucBin2}
\\
\sum_{s=t-\text{UT}_i+1}^{t}y_{s,i}&\leq x_{t,i} &&\text{\scriptsize $t\in\{\text{UT}_i,\dots, \mathcal{T}\}$}\label{eq:ucMinUp}
\\
\sum_{s=t-\text{DT}_i+1}^{t}z_{s,i}&\leq 1-x_{t,i}&& \text{\scriptsize $t\in\{\text{DT}_i,\dots, \mathcal{T}\}$}\label{eq:ucMinDown}
\\
p_{t,i}&\geq \underline{P}_i x_{t,i} &&\text{\scriptsize $t\in \mathcal{T},\;i\in\mathcal{I}$}\label{eq:ucMinGen}
\\
p_{t,i}+\rho_{t,i}&\leq \overline{P}_i x_{t,i} &&\text{\scriptsize $t\in \mathcal{T},\;i\in\mathcal{I}$}\label{eq:ucMaxGen}
\\
p_{t-1,i}-p_{t,i}&\leq \underline{\mathcal{R}}_i &&\text{\scriptsize $t\in \mathcal{T},\;i\in\mathcal{I}$}\label{eq:ucRampDown}
\\
p_{t,i}-p_{t-1,i}&\leq \overline{\mathcal{R}}_i &&\text{\scriptsize $t\in \mathcal{T},\;i\in\mathcal{I}$}\label{eq:ucRampUp}
\\
\sum_{i\in\mathcal{I}}\big(p_{t,i}\big)+p^{\text{\tiny RES}}_t&={P_d}_t &&\text{\scriptsize $t\in \mathcal{T}$}\label{eq:ucPowerBalance}\\
\sum\limits_{i \in \mathcal{I}, i\neq \ell}\rho_{t,i}&\geq p_\ell&&\text{\scriptsize $t\in \mathcal{T},\;\forall \ell$}\label{eq:ucResReq}
\end{align}
\end{subequations}

Where \cref{eq:ucObj} is the objective function, which is subjected to \cref{eq:ucBin1} to~\cref{eq:ucResReq}. $\text{suc}(\cdot)$ and $\text{gc}(\cdot)$ are the start-up and generation cost functions in euros, respectively. $\text{gc}(\cdot)$ is the quadratic cost function, which is piecewise linearized and then utilized in the \gls{UC} problem. $x$, $y$, and $z$ are binary variables of commitment, start-up, and shut-down, respectively. $p$, $\rho$, $p^{\text{\tiny RES}}$ are the thermal, reserve, and \gls{RES} power generation variables in megawatts. $t$ is the index for time intervals. $i$ is the index of generators. $\overline{P}$ and $\underline{P}$ are the maximum and minimum power output of generators, respectively. $\overline{\mathcal{R}}$ and $\underline{\mathcal{R}}$ are ramp-up and ramp-down limitation of generators, respectively. ${P_d}$ is the demand.

The aim is to solve \cref{eq:ucObj} subject to \cref{eq:ucBin1,eq:ucBin2,eq:ucMinUp,eq:ucMinDown,eq:ucMinGen,eq:ucMaxGen,eq:ucRampDown,eq:ucRampUp,eq:ucPowerBalance,eq:ucResReq}.
\Cref{eq:ucBin1,eq:ucBin2} represent the binary logic of the \gls{UC} problem. \Cref{eq:ucMinUp,eq:ucMinDown} are the minimum up-time and minimum downtime constraints of the units. \Cref{eq:ucMinGen} is the minimum power generation constraint. \Cref{eq:ucMaxGen} is the maximum power generation constraint and states that the summation of power generation and power reserve of every online unit, should be less than the maximum output of the unit. \Cref{eq:ucRampDown,eq:ucRampUp} are ramp-down and ramp-up constraints. \Cref{eq:ucPowerBalance} is the power balance equation. \Cref{eq:ucResReq} is the spinning reserve constraint and makes sure that there is enough reserve to compensate for the active power disturbance in case of loss of the generating unit $\ell$.

To include the constraint on the duration of frequency deviation under the threshold, the linearized version of \cref{eq:swing2_ber,eq:second_order_r_ber} should be added to \cref{eqs:uc}, and then by imposing \cref{eq:a_constraint} the duration of $h_\text{\tiny th}$ will be restricted.

\section{Results Regarding Solving the UC problem}\label{sec:result}

This section presents the results of applying the second-order frequency dynamics formulation within the \gls{UC} problem to a Spanish island system. Details of the case study are provided in Appendix~VII. As illustrated in \cref{fig:durationVsArea}, we regulate the parameter $h_<$ by imposing lower bounds on $\underline{A}$. The base case, which serves as a reference, imposes no constraint on $\underline{A}$ and corresponds to the standard \gls{UC} problem without frequency considerations. In addition, we include a scenario that augments the standard \gls{UC} with the frequency nadir constraint defined in \cref{eq:nadir}. To assess the impact of the proposed \gls{FCUC} model, two further scenarios are evaluated by setting $\underline{A} > 0$ and $\underline{A} > 1$, respectively. All optimization problems are solved using the Gurobi mixed-integer programming solver~\cite{gurobi}.
These cases are compared in terms of their operation cost in \cref{tab:sum_res}.
\begin{table}[!htbp]
    \centering
    \caption{Operation cost of the considered cases.}
    \label{tab:sum_res}
    \begin{tabular}{ccccc}
    \toprule
         & operation cost & rows & columns & nonzeros \\ \midrule
       \multirow{ 2}{*}{base case}  & \multirow{ 2}{*}{$69.84$ k€} & \multirow{ 2}{*}{\num{3966}} & \num{1086} continuous & \multirow{ 2}{*}{\num{15008}}\\
       &&&\num{2100} integer&\\
       
       \multirow{ 2}{*}{with nadir}  & \multirow{ 2}{*}{$70.60$ k€} & \multirow{ 2}{*}{\num{250809}} & \num{107348} continuous & \multirow{ 2}{*}{\num{848539}}\\
       &&&\num{1603} integer&\\
       
       $\underline{A} > 0$ & $70.18$ k€ & \multirow{ 2}{*}{\num{251379}} & \num{107348} continuous  & \multirow{ 2}{*}{\num{855994}}\\
       $\underline{A} > 1$ & $70.44$ k€ & & \num{1603} integer&\\\bottomrule
    \end{tabular}
\end{table}
As expected, the operational cost is higher for the \gls{FCUC} formulation. However, as will be demonstrated in the following sections, this increase in cost is accompanied by a significant improvement in frequency response, highlighting the trade-off between economic efficiency and system stability.
 
\Cref{tab:sum_res} presents the number of rows (constraints), columns (variables), and nonzeros (nonzero elements in the coefficient matrix of the constraints) after the solver, in this case, Gurobi, has presolved the problem. Due to the inclusion of frequency-related constraints, the size of the \gls{FCUC} problem is significantly larger than that of the base case. However, both cases with $\underline{A} > 0$ and $\underline{A} > 1$ result in the same problem size.

Despite this increase in problem size, the \gls{FCUC} formulation remains smaller than many optimization problems routinely solved in power systems, such as the \gls{UC} for large-scale systems. This is primarily because the case study under consideration is inherently small. For large systems, solving \gls{FCUC} problems with the level of precision proposed in this paper is unnecessary. Frequency stability poses a greater challenge in small systems, such as the one studied here, necessitating a more detailed frequency dynamic model, as presented in this work. Therefore, the scalability of the proposed formulation is not a critical concern.

\subsection{Base case}

\begin{figure}[!htbp]
    \centering
    \includegraphics[width=0.8\linewidth]{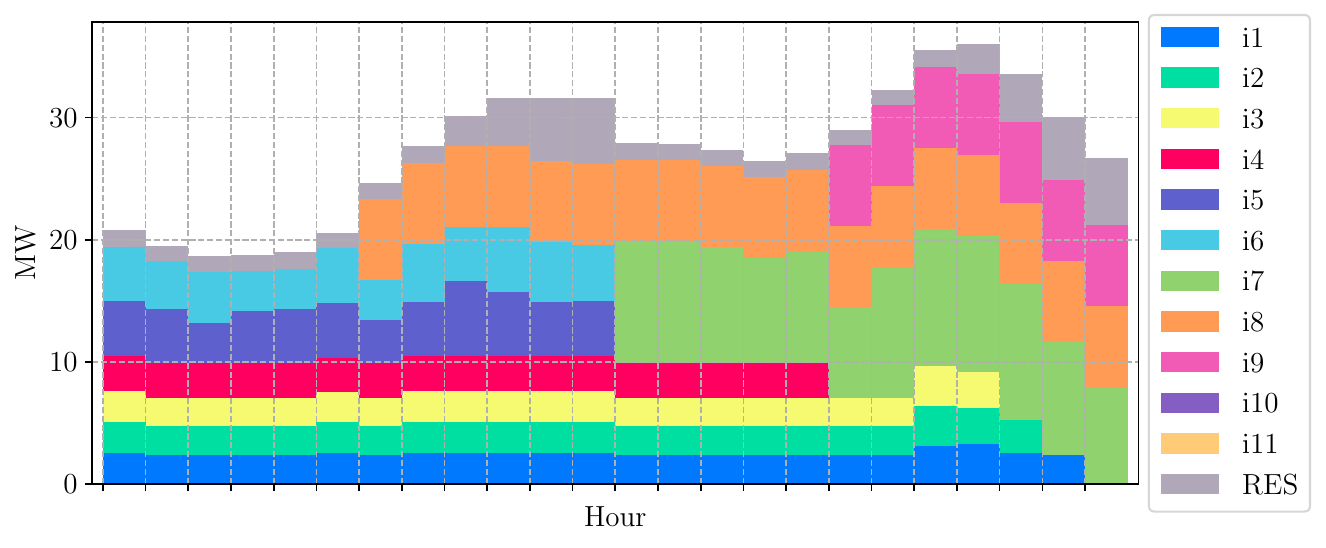}
    \caption{Generation schedule of the base case.}
    \label{fig:gen_A1000}
\end{figure}

This case corresponds to the \gls{UC} problem without frequency constraints (\cref{eq:ucObj} to \cref{eq:ucResReq}). The generation schedule of the units obtained after solving this case is presented in \cref{fig:gen_A1000}.

Examining the $\underline{A}$ values for each outage, we observe that in some incidents, $\underline{A}$ takes significantly negative values. This indicates that the frequency will remain below $\Delta f_{\text{th}}$ for an extended period.
\begin{figure}[!htbp]
    \centering
    \includegraphics[width=0.8\linewidth]{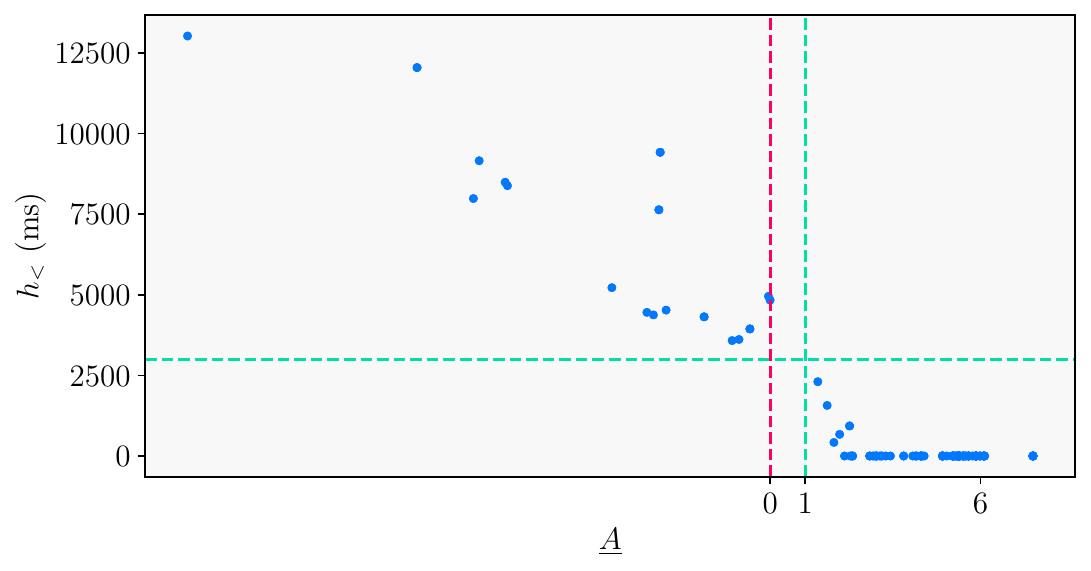}
    \caption{The duration of frequency under threshold ($h_<$) as a function of $\underline{A}$ for every outage of the base case.}
    \label{fig:Amin_1000}
\end{figure}
\Cref{fig:Amin_1000} illustrates the duration of frequency excursions below the threshold for each outage ($h_<$) as a function of $\underline{A}$. The values of $\underline{A}$ are obtained using an exact second-order differential model after solving the \gls{UC} problem. Many outages result in poor frequency response, where the frequency drops below the threshold and remains there for several seconds. Such incidents will inevitably trigger under-frequency load shedding relays, indicating that load shedding is necessary to maintain system stability.

\subsection{With nadir constraint}

\begin{figure}[!htbp]
    \centering
    \includegraphics[width=0.8\linewidth]{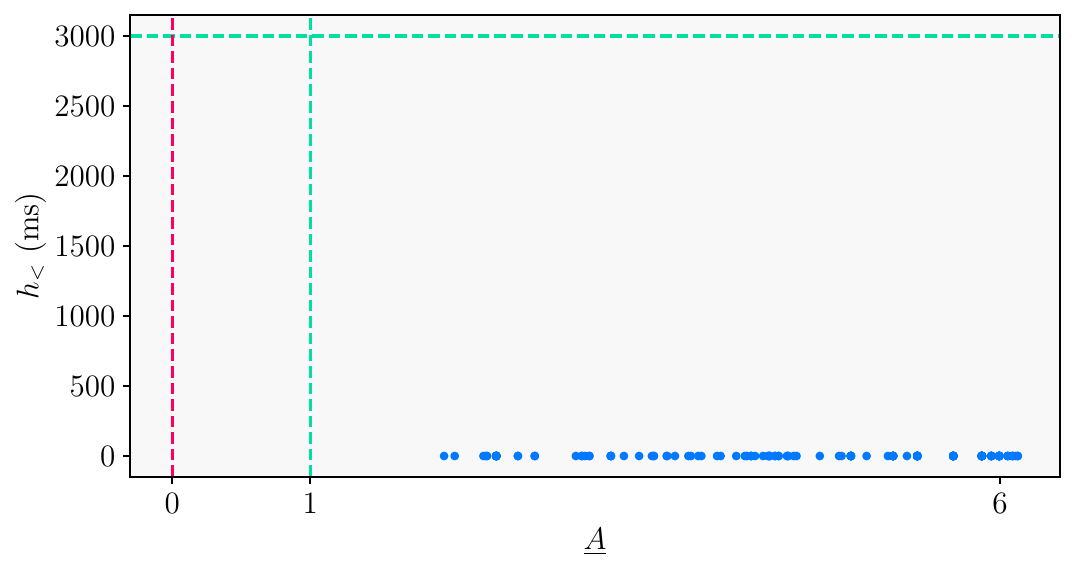}
    \caption{Duration of frequency falling below the threshold ($h_<$) as a function of $\underline{A}$ for each outage in the case with nadir constraint.}
    \label{fig:Amin_nadir}
\end{figure}

Enforcing the nadir constraint in \cref{eq:nadir} ensures that the frequency response never falls below the specified threshold. This approach is more conservative than allowing temporary excursions below the threshold for short durations. As shown in \cref{tab:sum_res}, this added conservatism results in higher operational costs, making this scenario the most expensive among those evaluated. Consequently, it is expected that $\underline{A}$ remains zero for all outages under this constraint. This behavior is confirmed in \cref{fig:Amin_nadir}.

\subsection{$\underline{A} > 0$}\label{sec:res_A0}

In this case, the constraint $\underline{A} > 0$ is enforced. As shown in \cref{fig:durationVsArea}, this condition should effectively prevent incidents where the frequency drops below the threshold for more than three seconds.
\begin{figure}[!htbp]
    \centering
    \includegraphics[width=0.8\linewidth]{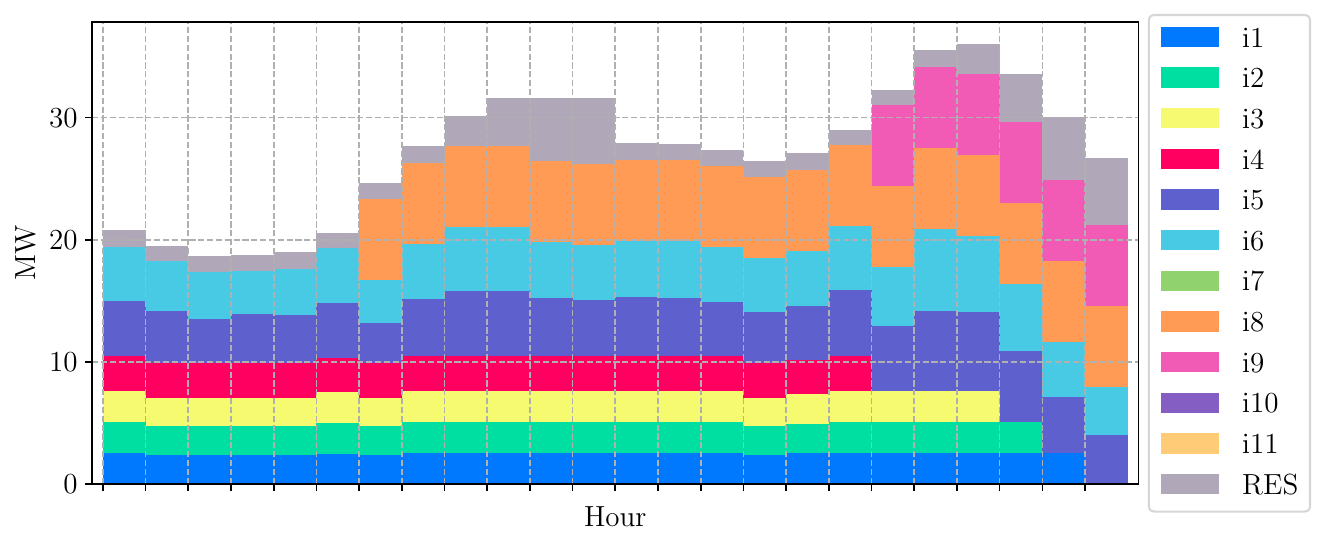}
    \caption{Generation schedule for the case with $\underline{A} > 0$.}
    \label{fig:gen_A0}
\end{figure}
As illustrated in \cref{fig:gen_A0}, the maximum outage during peak hours is reduced compared to \cref{fig:gen_A1000}. More generating units with lower generation levels are now scheduled during peak hours to mitigate the occurrence of large outages.
\begin{figure}[!htbp]
    \centering
    \includegraphics[width=0.8\linewidth]{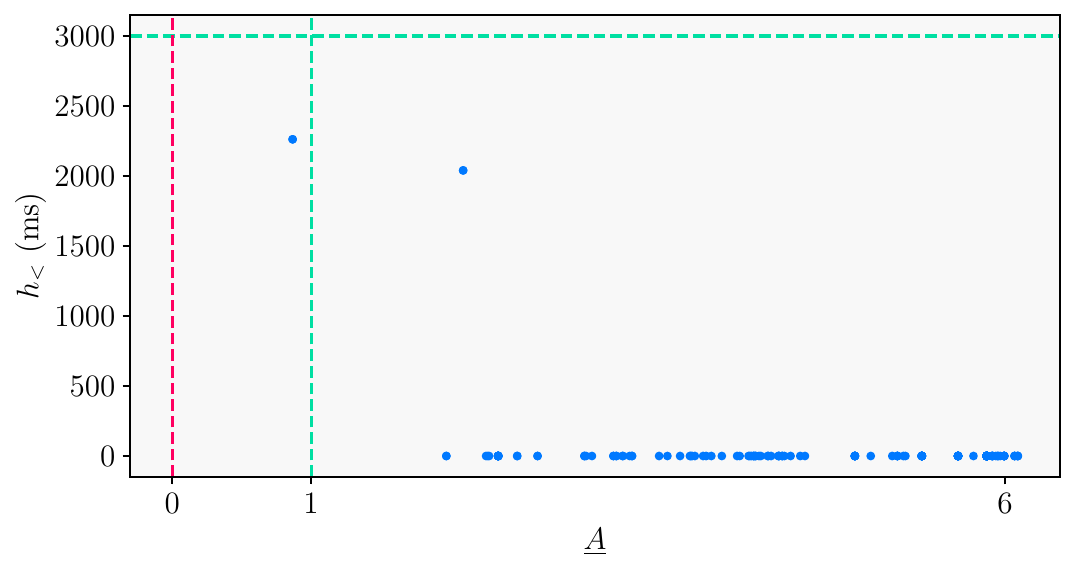}
    \caption{Duration of frequency falling below the threshold ($h_<$) as a function of $\underline{A}$ for each outage in the case with $\underline{A} > 0$.}
    \label{fig:Amin_0}
\end{figure}
As confirmed by \cref{fig:Amin_0}, enforcing $\underline{A} > 0$ is sufficient to prevent any outage that would cause the frequency to remain below the threshold for more than three seconds. Although the frequency dips below the threshold in some instances, it recovers within the intended duration. This confirms that the data-driven approach proposed in \cref{sec:dataDrivenApproach} operates as intended.

\subsection{$\underline{A} > 1$}

This case is more restrictive compared to \cref{sec:res_A0}. It keeps $\underline{A}$ always above 1 (see \cref{fig:durationVsArea}). 
\begin{figure}[!htbp]
    \centering
    \includegraphics[width=0.8\linewidth]{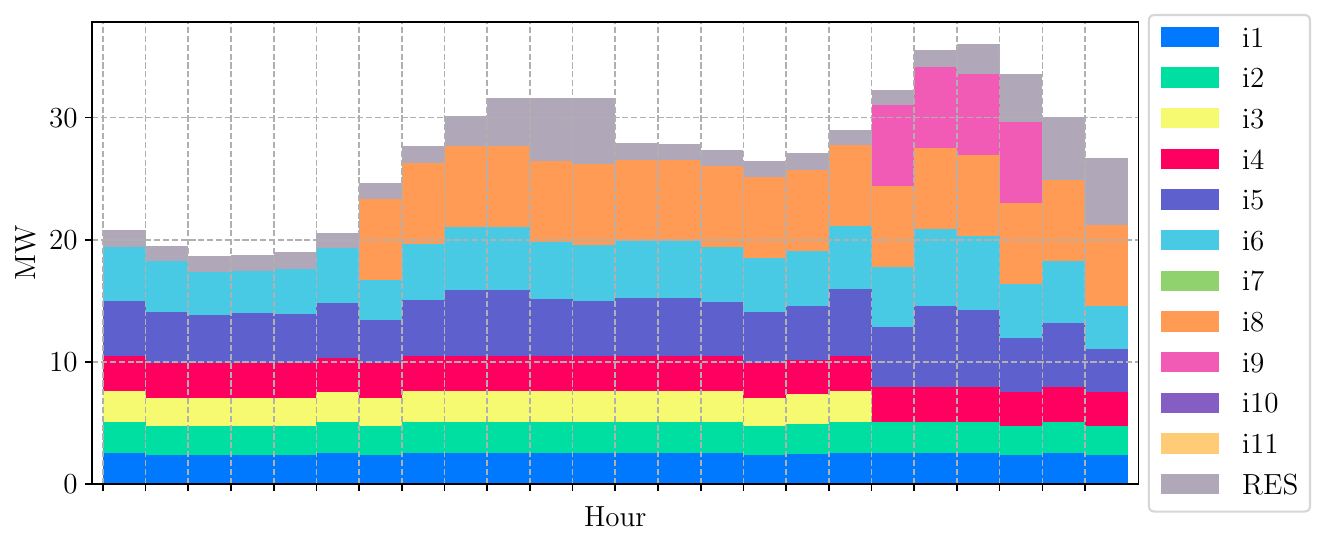}
    \caption{Generation schedule for the case with $\underline{A} > 1$.}
    \label{fig:gen_A1}
\end{figure}
The scheduled generation of the units is shown in \cref{fig:gen_A1}.
\begin{figure}[!htbp]
    \centering
    \includegraphics[width=0.8\linewidth]{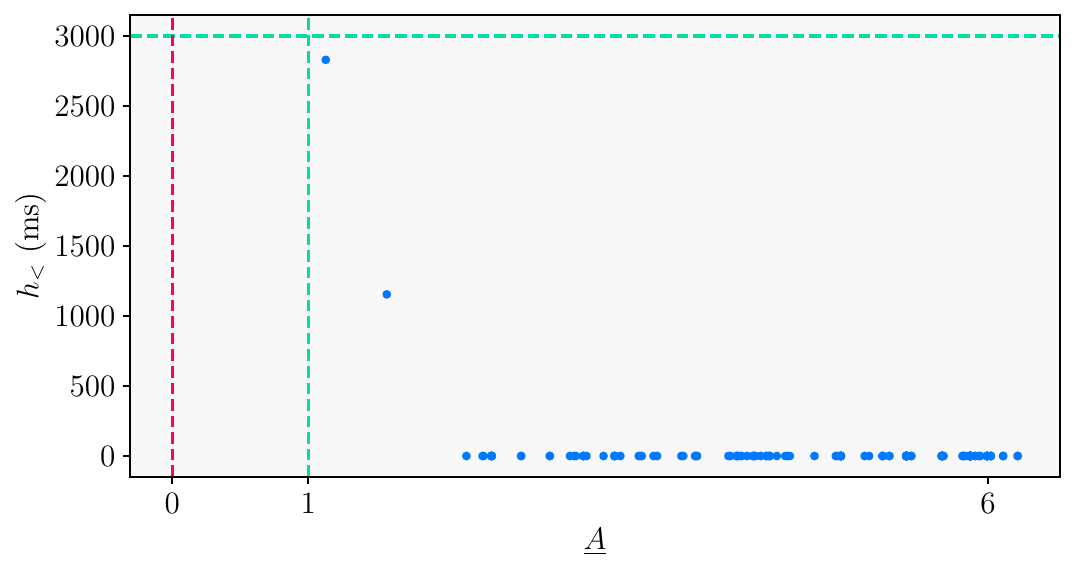}
    \caption{Duration of frequency falling below the threshold ($h_<$) as a function of $\underline{A}$ for each outage in the case with $\underline{A} > 1$.}
    \label{fig:Amin_1}
\end{figure}
And in \cref{fig:Amin_1} $\underline{A}$ of each outage is depicted. \Cref{eq:a_constraint} has been able to keep $\underline{A}$ above 1 for every single outage.

\section{Conclusions}\label{sec:conc}

This paper presented a new method for modeling frequency dynamics within the unit commitment problem using general-order ordinary differential equations solved with Bernstein polynomial approximation. Unlike prior works that rely on linear assumptions or scalar frequency metrics, our approach models full second-order generator dynamics, and through Bernstein operational matrices, derives a tight and tractable MILP reformulation. A key innovation is the integration of a duration-based frequency constraint, targeting the time that system frequency remains below critical thresholds—an aspect overlooked in traditional \gls{FCUC} models. This constraint better aligns with real-world relay behavior and equipment vulnerability.

We further proposed a data-driven method to quantify the duration of frequency excursions using the integral of the frequency response approximation, circumventing the computational cost of binary encodings. Case studies on a Spanish island system illustrated the effectiveness of our method, showing that it achieves significant improvements in frequency security while preserving computational feasibility and operational efficiency. The approach is especially well-suited for low-inertia or islanded systems, where frequency stability is a critical constraint.

\section*{Appendix I: Available Reserve Limitation}

In reality, the capacity of units is limited. They can increase their output, until they reach their maximum capacity. 
\begin{figure}[!htbp]
    \centering
    \includegraphics[width=0.7\linewidth]{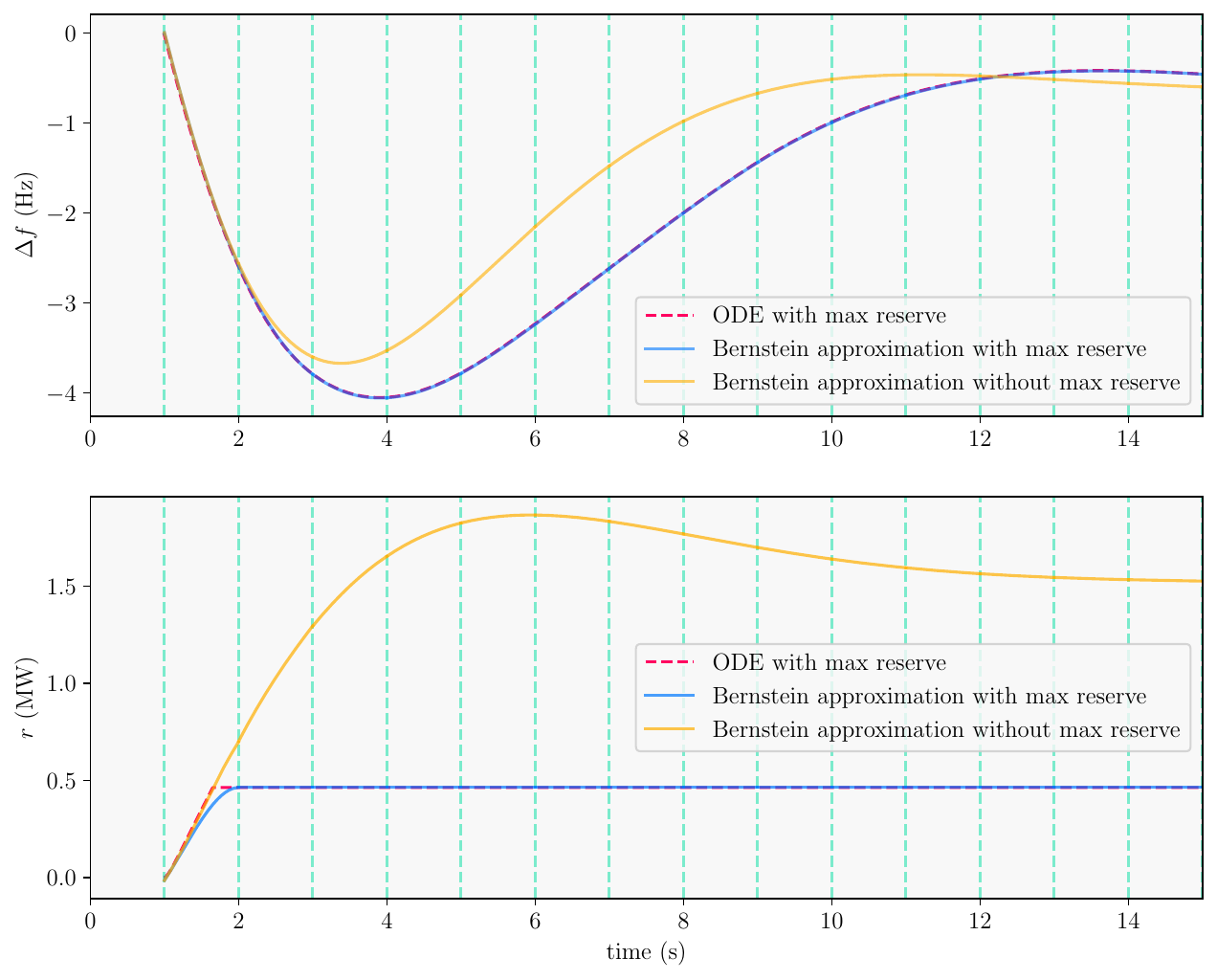}
    \caption{Approximating frequency response while considering the $\min$ function in \cref{eq:swing} Vs. while not considering the $\min$ function in \cref{eq:swing}.}
    \label{fig:maxRes}
\end{figure}
\Cref{fig:maxRes} compares the frequency response when reserve limit is considered versus when it's not. The dashed lines are exact numerical solutions computed with the SciPy function odeint \cite{ahnert2011odeint}. When the reserve limitation is not considered, the responded reserve is increased freely, leading to a big underestimation of frequency decay.

\section*{Appendix II: Bernstein Polynomial Intervals}

Reducing the number of intervals while maintaining similar accuracy is therefore advantageous. 
\begin{figure}[!htbp]
    \centering
    \includegraphics[width=0.7\linewidth]{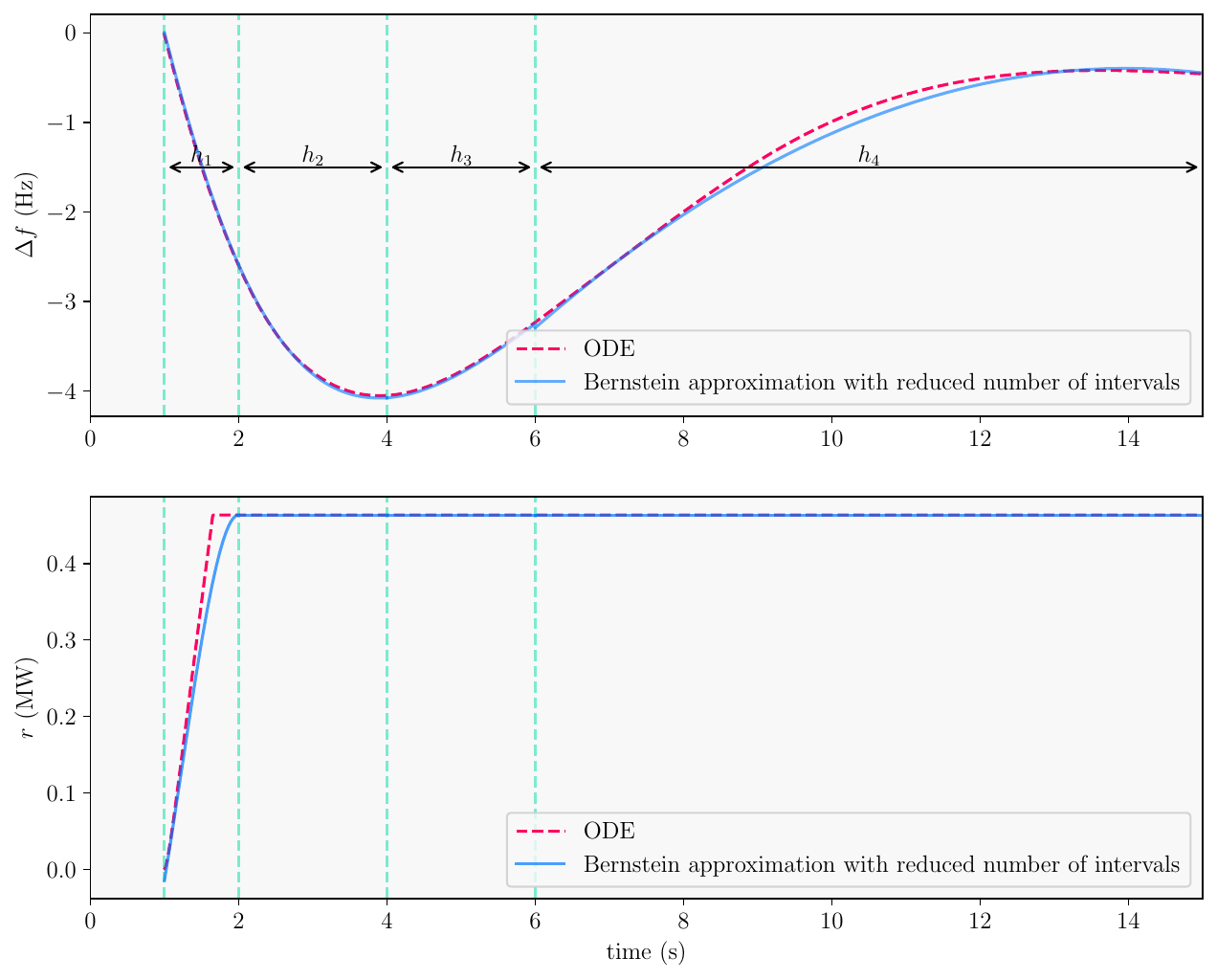}
    \caption{Approximating the second order model frequency response, with reduced number of intervals.}
    \label{fig:lessIntervals}
\end{figure}
\Cref{fig:lessIntervals} shows that the approximation has acceptable accuracy with much less intervals. Only one interval is used from 6 to 15 seconds, because not much accuracy is required for this part of the frequency response. 

\section*{Appendix III: Order of Bernstein Polynomials}

\begin{figure}[!htbp]
    \centering
    \includegraphics[width=0.7\linewidth]{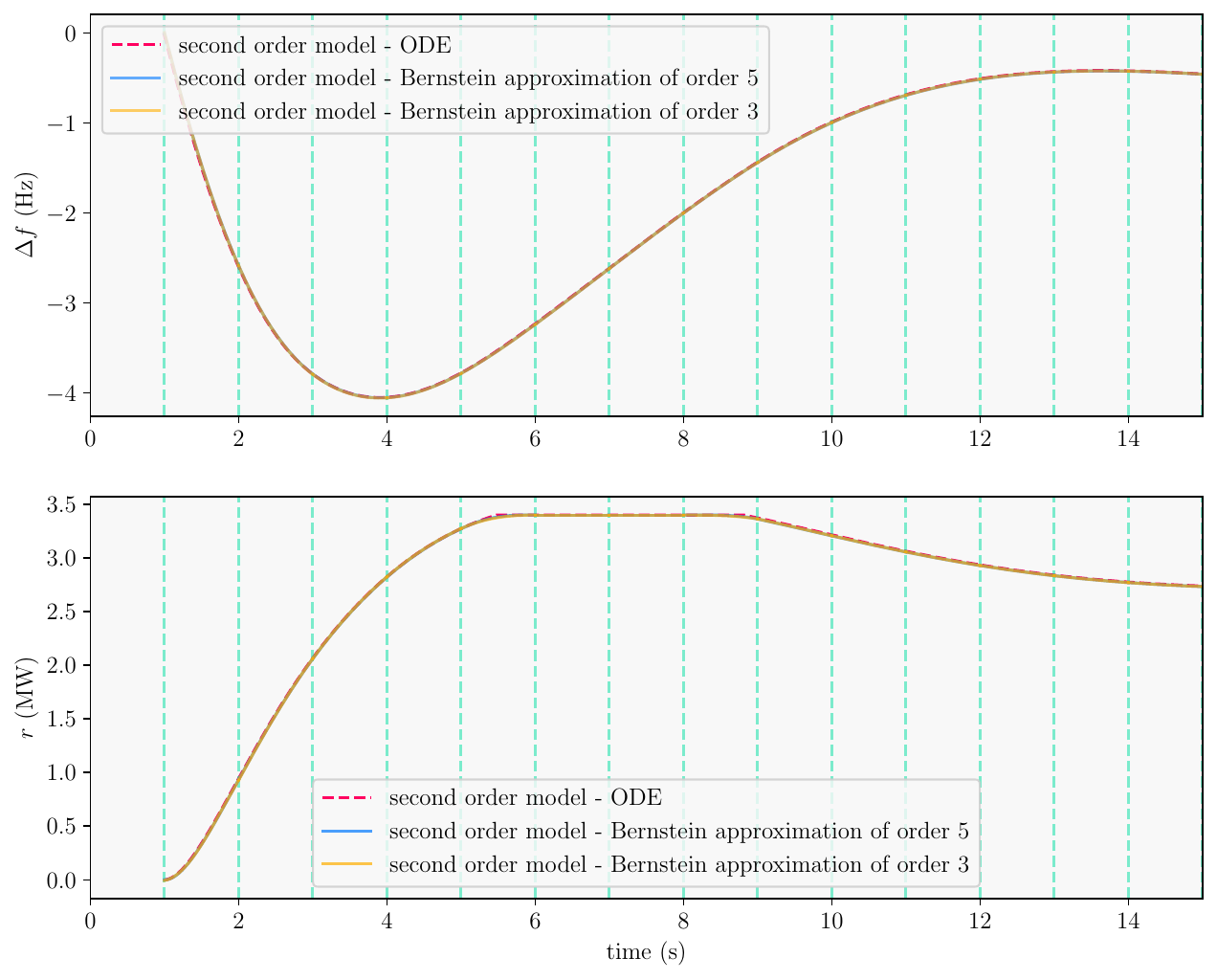}
    \caption{Bernstein polynomials of order 3 Vs. Bernstein polynomials of order 5. The only difference is the order of polynomials, while the length is kept the same ($h_\tau=1$ for every $\tau$).}
    \label{fig:bercomp}
\end{figure}

Arguably higher order polynomials are able to approximate the original function more accurately. \Cref{fig:bercomp} compares the accuracy of polynomials of order 3 and 5 with the exact solution of ODE.
It's evident from this example that polynomials of order 3 and 5 are both able to accurately follow the exact solution. In case of similar accuracy, we'd prefer to use polynomials of lower order, as that will reduce the complexity.

\section*{Appendix IV: First order Vs. Second Order Model}

\begin{figure}[!htbp]
    \centering
    \includegraphics[width=0.7\linewidth]{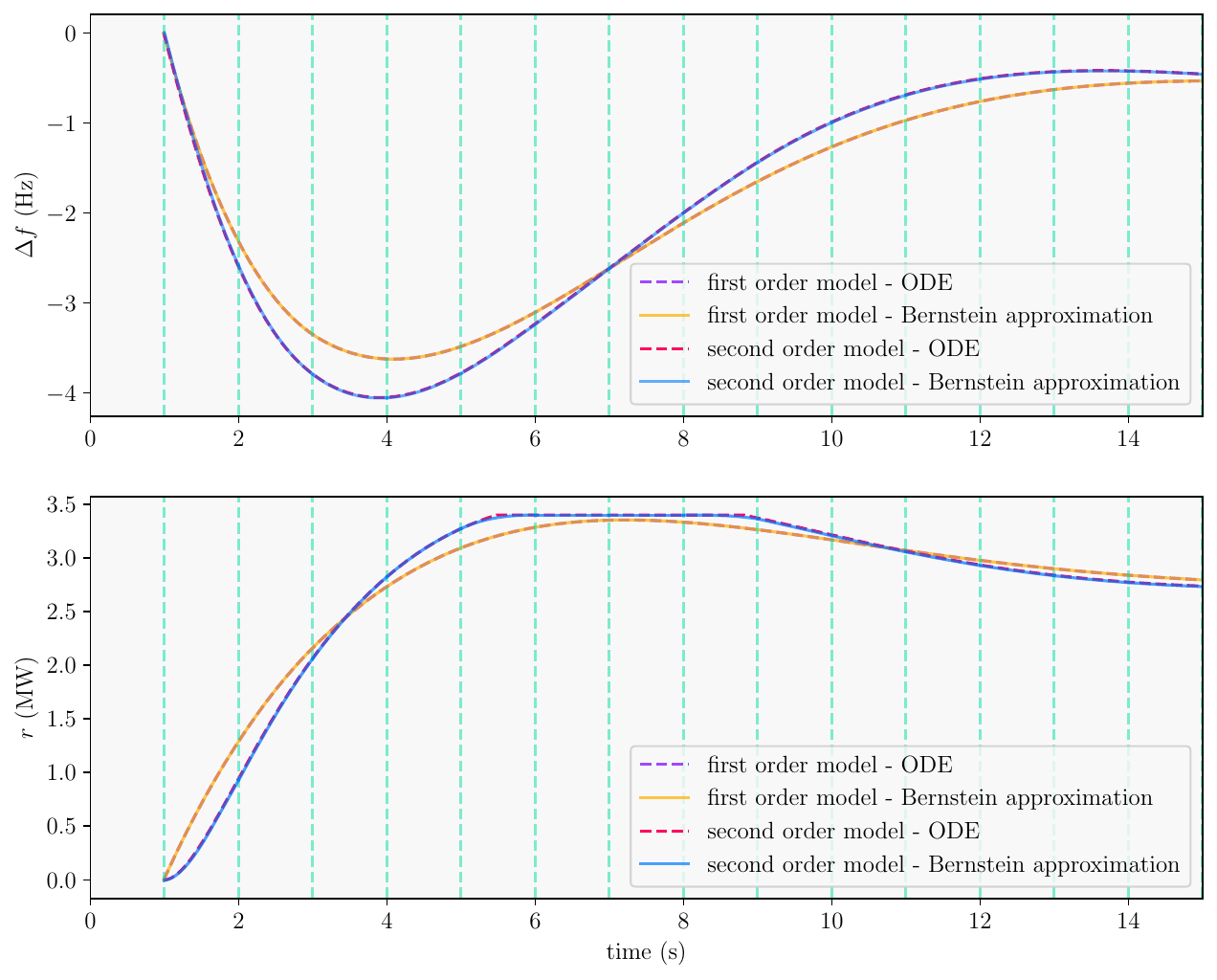}
    \caption{First order Vs. Second Order Model. The Bernstein approximation is done with polynomials of order 3 ($n=3$) and the length on 1 second ($h_\tau=1$ for every $\tau$).}
    \label{fig:firstVssecond}
\end{figure}

From \cref{fig:firstVssecond} it's evident that there is meaningful difference between the first order (\cref{sec:FOM}) and the second order (\cref{sec:SOM}) model.
The first order model tends to underestimate the frequency decay, leading to a lower frequency nadir deviation. 

\section*{Appendix V: Binary into Variable Linearization}\label{sec:appI}

There are some binary into variable non-linearities in the formulation so far that should be linearized so that we can use the formulation in a \gls{MILP} problem. Each $xc$, where $x$ is a binary variable and $c$ is a continuous variable, can be linearized with the following set of constraints,
\begin{subequations} \label{equ_12}
    \begin{align}
        a&\leq Mx\\
        a&\leq c\\
        a&\geq c-M(1-x)
    \end{align}
\end{subequations}
Where $a$ serves as an auxiliary variable for storing $xc$, while $M$ represents a sufficiently large number.

A second option is to treat $x$ as binary activation variable and to impose these logical terms,
\begin{equation}
    \begin{cases}
        x=0 \rightarrow a=0 \\
        x=1 \rightarrow a=c
    \end{cases}
\end{equation}
These terms are called indicator constraints and are readily accepted by modern MILP solvers like Gurobi and CPLEX. 

\section*{Appendix VI: Min(.) Term}\label{sec:appII}

The \texttt{min(.)} term in \cref{eq:second_order,eq:second_order_r_ber} is nonlinear. It can be linearized with a series of constraints. We enforce $r^{\text{\tiny min}}_i=\min \big(r_i,\rho_i\big)$ by,
\begin{subequations}
\begin{align}
    r^{\text{\tiny min}}_i &\geq \rho_i-M(1-z_i)\\
    r^{\text{\tiny min}}_i &\leq \rho_i\\
    r^{\text{\tiny min}}_i &\geq r_i-Mz_i\\
    r^{\text{\tiny min}}_i &\leq r_i\\
    r^{\text{\tiny min}}_i &\leq \rho_i + Mz_i\\
    \rho_i-r_i &\leq M(1-z_i)   
\end{align}
Again $z_i$ can be treated as a binary activation variable to avoid defining a value for $M$.

\end{subequations}

\section*{Appendix VII: Data of Generators}\label{sec:appGD}
Generator's data for the case study is presented in \cref{tab:data}.
\begin{table}[!htbp]
    \centering
    \caption{Data of the generators}
    \begin{tabular}{c|cccccc}
        i & $p_i$ [MW] & $p^{\max}_i$ [MW] & $H_i$ [s] & $K_i$ [-] & $T_i$ [-] & $b_i$ [-]  \\\hline
        1 & 3.36 & 3.82 & 1.75 & 20 & 17.26 & 1.82   \\
        2 & 0 &  3.82 & 1.75 & 20 & 17.26 & 1.82 \\
        3 & 0 & 3.82 & 1.75 & 20 & 17.26 & 1.82 \\
        4 & 2.82 & 4.3 & 1.73 & 20 & 17.06 & 1.80  \\
        5 & 3.3 & 6.7 & 2.16 & 20 & 24.63 & 3.16  \\
        6 & 0 &  6.7 & 1.88 & 20 & 18.79 & 2.05 \\
        7 & 0 & 11.2 & 2.10 & 20 & 24.93 & 3.21  \\
        8 & 6 & 11.5 & 2.10 & 20 & 24.93 & 3.21  \\
        9 & 9 & 11.5 & 2.10 & 20 & 24.93 & 3.21  \\
        10 & 0 & 11.5 & 2.10 & 20 & 24.93  & 3.21  \\
        11 & 0 & 21 & 6.5 & 21.25 & 4.43 & 0.83  \\
    \end{tabular}
    \label{tab:data}
\end{table}
The column $p_i$ [MW] is an arbitrary operating point. The outage of $i=9$ is showcased in other appendices in various situations.

\bibliographystyle{IEEEtran}
\bibliography{bib}\label{references}






\end{document}